**Field Orientation Dependent Magnetic Phases In Weyl Semimetal $Co_3Sn_2S_2$**

Samuel E. Pate[1,2], Bin Wang[3], Bing Shen[3], J. Samuel Jiang[1], Ulrich Welp[1], Wai-Kwong Kwok[1], Jing Xu[4], Kezhen Li[1,2], Ralu Divan[4], and Zhi-Li Xiao[1,2,*]

[1]Materials Science Division, Argonne National Laboratory, Argonne, Illinois 60439, United States

[2]Department of Physics, Northern Illinois University, DeKalb, Illinois 60115, United States

[3]School of Physics, Sun Yat-sen University, Guangzhou 510275, China

[4]Center for Nanoscale Materials, Argonne National Laboratory, Argonne, Illinois 60439, United States

Magnetism plays a key role in the emergence of topological phenomena in the Weyl semimetal $Co_3Sn_2S_2$, which exhibits a ferromagnetic (FM) interactions along the $c$-axis of the crystal and an antiferromagnetic (AFM) interactions within the $ab$ plane. Extensive studies on the temperature dependence of the magnetism with the magnetic field along the $c$-axis have uncovered a number of magnetic phases. Currently, the nature and origins of the reported magnetic phases are under debate. Here we report on magnetic field orientation effects on the magnetism in $Co_3Sn_2S_2$. The shape of the hysteresis loop of the Hall resistance at a fixed temperature is found to change from rectangular to bow-tie-like as the magnetic field is tilted from the $c$-axis towards the $ab$ plane, resembling that reported for magnetic fields along the $c$-axis as the temperature approaches the Curie temperature from below. Unlike their temperature-dependent counterparts, the newly observed bow-tie-like hysteresis loops show exchange bias. Our results showcase the contribution of the in-plane AFM interactions to the magnetism in $Co_3Sn_2S_2$ and demonstrate a new way to tune its magnetic phases. They also shed light on the temperature-dependent magnetic phases occurring in the magnetic field along the $c$-axis of the crystal.

*Corresponding author, xiao@anl.gov or zxiao@niu.edu

Recently, $Co_3Sn_2S_2$ was discovered to be a Weyl semimetal [1-4] with topologically nontrivial behavior such as giant anomalous Hall effect [1,2] and Hall angle [1,5], chiral-anomaly-induced negative magnetoresistance [1], large magneto-optical response [6], and strong spin polarization [7,8]. It exhibits magnetic order below the Curie temperature $T_c$ [1,2,6], arising from the $d$ orbitals of Co ions arranged in a kagomé lattice in the $ab$ plane [9]. First-principle calculations [10] reveal ferromagnetism (FM) with Co magnetic moments aligned parallel to the $c$-axis. Experiments at low temperatures confirm that $c$-axis of the crystal is the easy axis of magnetization and a field of $H = 23$ Tesla is required to fully orientate the Co magnetic moment parallel to the kagomé plane at $T = 2$ K [11].

As the first confirmed intrinsic time reversal symmetry-breaking Weyl semimetal, $Co_3Sn_2S_2$ has launched extensive investigations on its magnetic order [12-27] that is strongly correlated with the bandstructure and topological semimetallic states [14-16,19,28,29], leading to the discovery of intriguing magnetic phases [12-15,18-20,22-27]. For example, temperature dependent ac magnetic susceptibility measurements uncovered an anomalous magnetic phase (denoted as "A" phase) at low fields near $T_c$ [12,19]. High-resolution muon spin-rotation (μSR) experiments indicate that $Co_3Sn_2S_2$ crystals exhibit a $c$-axis FM ground state at low temperatures. When the temperature is increased towards $T_c$, the volume of the FM component is found to decrease while an in-plane antiferromagnetic (AFM) phase emerges and becomes dominant at temperatures near $T_c$. The magnetic competition between these two phases is found to be highly tunable by applying pressure [13,14] or doping [21,22]. New magnetic phases have also been suggested from the measurements on the magnetic field dependence of the magnetization $M(H)$ and anomalous Hall effect $R_{xy}(H)$, where a change from rectangular to bow-tie-like structure of the $M(H)$ and $R_{xy}(H)$ hysteresis loops was observed and attributed to a freezing transition of the spin glass state arising from the strong magnetic frustration in the $ab$ plane [15].

Currently, experiments revealing possible magnetic phases focus on thermal effects, with magnetic fields applied along the *c*-axis of the crystal. Although extensive investigations have been conducted, the nature and origins of these possible magnetic phases, particularly the contribution of the in-plane AFM structures, have been widely debated [14-29]. In this Letter, we investigate the effects of magnetic field orientation on magnetic phases by tilting the magnetic field from the *c*-axis towards the *ab* plane. We measure $R_{xy}(H)$ hysteresis loops at a fixed temperature while changing the direction of the magnetic field. We find that the shape of the hysteresis loops can change from rectangular to bow-tie-like when the magnetic field is rotated from the *c*-axis towards the *ab* plane. Remarkably, this behavior is similar to that observed for magnetic field along the *c*-axis [15] when the temperature is increased towards $T_c$. The shape variation in the $R_{xy}(H)$ hysteresis loops driven by the magnetic field orientation is observed at temperatures down to the lowest experimentally available temperature ($T = 3$ K) while occurring only at a temperature-insensitive and narrow angle range (< 6°) from the *ab* plane. Together with the accompanying exchange bias, the bow-tie-like $R_{xy}(H)$ hysteresis loops observed in the vicinity of the *ab* plane highlight the role of AFM interaction in the magnetism in $Co_3Sn_2S_2$. Our results also provide evidence for the contribution of the AFM structure to the appearance of previously reported temperature-dependent magnetic phases and demonstrate a new way to tune the magnetic phases of this Weyl semimetal.

Our measurements were conducted on two crystals grown by a self-flux method [1]. They show consistent results and data presented here are from one of them. The $R_{xy}(H)$ hysteresis loops were obtained using the Electrical Transport Option (ETO) of a Quantum Design PPMS®. Electrical leads were gold wires arranged in a Hall bar geometry, glued to the crystal using silver epoxy H20E. A low-frequency (21.36 Hz) ac current of 1 mA was applied in the *ab* plane and the angular dependencies of the resistance were obtained by placing the sample on a precision, stepper-controlled rotator with an angular resolution of 0.005°. The magnetic field was rotated in the plane perpendicular to the current. The magnetic field orientation is represented by $\theta$, which is defined as the angle between the applied magnetic field and the *c*-axis. Experimentally, the orientation of *ab* plane with respect to the magnetic field is determined by measuring the angle dependence of the Hall resistance $R_{xy}$ at $H = \pm 7$ T. The crossing point of the two curves for $H = \pm 7$ T is defined as $\theta = 90°$, corresponding to **H**∥*ab*. To avoid contribution of magnetic moment to the anisotropy of $R_{xy}$, we conducted measurements in the paramagnetic state, i.e., at $T = 300$ K and the results are presented in Fig.S1 in the Supplementary Material [30].

As indicated by the resistance versus temperature curve presented in Fig.S2 in the Supplemental Material [30] for zero-field cooling, our crystal has a Curie temperature of $T_c \approx 173$ K, consistent with those in the literature, ranging from $T_c \approx 172$ K to 177 K [1,2,9,14-17]. Figure 1 shows $R_{xy}(H)$ loops obtained at various temperatures for magnetic fields aligned along the *c*-axis, which exhibit similar temperature-induced shape changes as those reported in the literature [15]. More specifically, the loop at $T = 100$ K is rectangular with binary switchable magnetic moments, reflecting collective spin flips while the loops at $T = 145$ K and 151 K show bow-tie-like structures, as termed in Ref.15, with the former exhibiting 'triangular tails' and the latter showing 'double triangles'. More detailed measurements allowed us to identify a temperature of $T_A \approx 136$ K (see Fig.S3 and its caption in the Supplemental Material [30]) that clearly separates regimes where $R_{xy}(H)$ loops are rectangular (at $T < T_A$) and bow-tie-like (at $T > T_A$). This value of $T_A$ is nearly the same as those ($T_A = 135$ K [12] and $\approx 130$ K [19]) at which an anomaly occurs in the magnetic susceptibility. Moreover, it is about 10 K higher than the spin glass freezing temperature of $T_G \approx 125$ K [15]. In reported studies, the temperature above which anomalous magnetic phases were observed varies between 125 K and 135 K [6,12,18,19,23-26].

The bow-tie-like structure of the hysteresis loops in $Co_3Sn_2S_2$ at $T > T_A$ was understood as the interplay of two terms in the free energy in a minimal Landau model, one favoring a FM structure $M = \pm M_0$, and another favoring in-plane AFM structures $M = 0$ [15]. In other words, hysteresis loops with bow-tie-like structure indicate the coexistence of the out-of-plane FM order with in-plane AFM structure. Due to the absence of exchange bias (EB) expected for a mixed phase of FM and AFM orders, it was hypothesized [15] that the in-plane AFM structure is a soft or dynamic spin glass arising from the magnetic frustration in the kagomé lattice, which coexists but not interact with the FM structure. With decreasing temperature, the spin glass becomes frozen, leading to the occurrence of EB observed in the sample at $T < T_A$. That is, $T_A$ is the glass freezing transition temperature (thus was denoted as $T_G$ in [15]).

Since AFM structure was also detected in µSR measurements at $T > T_A$ [14], the above understanding [15] seems to be plausible. Indeed, $M(H)$ hysteresis loops with very similar bow-tie-like structures as those reported in Ref.15 and resembling our $R_{xy}(H)$ loops in Fig.1 were also observed in rare-earth orthoferrites [31-34], in which FM and AFM structures can coexist and the expected EB often does not occur either [32,34]. However, the bow-tie-like hysteresis loops without EB in rare-earth orthoferrites were explained by a reversible motion of a single magnetic domain wall in the sample, i.e., without involving the mixture of FM and AFM orders [32]. Furthermore, 'double triangles' were observed in hysteresis loops of materials without AFM structure [35,36]. Thus, further evidence is desirable to support the current understanding of the observed bow-tie-like structures in the hysteresis loops of $Co_3Sn_2S_2$.

Our key findings are displayed in Fig.2, which presents $R_{xy}(H)$ loops at $T < T_A$ and for magnetic fields orientated close to the *ab* plane. It demonstrates that $R_{xy}(H)$ loops can exhibit bow-tie-like structures at $T < T_A$ with concurrent EB, implying that antiferromagnetism may indeed contribute to the bow-tie-like structures in the $R_{xy}(H)$ loops at $T > T_A$, regardless of the EB's absence due to the temperature-induced softening of the AFM structure. Since $T$ (= 100 K) < $T_A$ (= 136 K), the $R_{xy}(H)$ loop is expected to be rectangular for $\boldsymbol{H}\|c$, i.e., at $\theta = 0°$, as shown in Fig.1(b). The rectangular shape stays unchanged even at an angle as large as $\theta = 84.5°$ (Fig.2(b)). However, a small tail occurs at the negative field side of the $R_{xy}(H)$ loop at $\theta = 85.0°$ (Fig.2(c)). When the magnetic field is further tilted towards the $ab$ plane, triangular tails appear on both sides of the loops as exhibited by those at $\theta = 86.5°$ (Fig.2(d)) and $86.8°$ (Fig.2(e)), followed by hysteresis loops with 'double triangles', as shown in Fig.2(f) and 2(g) for $\theta = 87.3°$ and $\theta = 87.5°$, respectively. The bow-tie-like structures eventually vanish when the magnetic field is within about 2° of the $ab$ plane (see Fig.2(h) for $\theta = 88.5°$). That is, magnetic phases similar to those observed at $\boldsymbol{H}\|c$ and $T > T_A$ (Fig.1 and Ref.15) can occur at a fixed temperature $T < T_A$ as the magnetic field tilts towards the $ab$ plane. Furthermore, unlike those for $\boldsymbol{H}\|c$ and $T > T_A$, the $R_{xy}(H)$ loops with bow-tie-like structures in Fig.2(c)-2(g) are asymmetric with respect to the zero field, i.e., the loops are exchange biased.

The evolution of the shape of the $R_{xy}(H)$ loops with the magnetic field orientation is further illustrated by the angle dependence of the switching fields, as presented in Fig.3. The switching fields $H_s^{np}$ and $H_s^{pn}$ are the values of the magnetic field prior to the abrupt changes in $R_{xy}$ when the magnetic field is swept from negative to positive and from positive to negative, respectively, as denoted in Fig.2(e). At $\theta \leq 82.5°$, $H_s^{pn}$ and $H_s^{np}$ monotonically increase as the magnetic field tilts away from the $c$-axis towards the $ab$ plane while having the same values but with opposite signs (Fig.3(a)), i.e., $|H_s^{pn}| = H_s^{np}$, reflecting the symmetry of the $R_{xy}(H)$ loops with respect to the zero field. At larger angles (see Fig.3(b) for an expanded view), the values of $H_s^{pn}$ and $H_s^{np}$ fluctuate and in most cases are not equal, indicating possible exchange bias. In particular, they start to decrease with increasing angle after a small tail occurs at the negative field side of $R_{xy}(H)$ loop (at $\theta > 85.0°$, Fig.2(c)). In a very narrow angle regime ($87° < \theta < 88°$) associated with occurrence of the 'double-triangles' in the $R_{xy}(H)$ loops (Fig.2(f) and 2(g)), $H_s^{pn}$ and $H_s^{np}$ change their signs and their values also increase rapidly.

Exchange bias is typically parameterized as $H_{EB} = (H_{+C}+H_{-C})/2$, where $H_{+C}$ and $H_{-C}$ represent the coercive fields for the positive and negative fields, respectively, reflecting that the loop is centered around a non-zero field $H_{EB}$ instead of $H = 0$ T [15,37]. While it works well for rectangular loops, such a definition of $H_{EB}$ does not appropriately reflect the asymmetry in the hysteresis loops with bow-tie-like structures (see Fig.2(e) and 2f)). In our analysis we replace $H_{+C}$ and $H_{-C}$ with $H_s^{np}$ and $H_s^{pn}$, respectively, i.e., $H_{EB} = (H_s^{np}+ H_s^{pn})/2$, which is equivalent to the conventional definition when the loops are rectangular. Figure 3(c) exhibits the angle dependence of $H_{EB}$ as error bars to the averaged switching field $H_s^{av} = (H_s^{np}- H_s^{pn})/2$. It shows that the loops with 'double-triangles' always have significant EBs while those with 'triangle-tails' can sometimes be symmetric, i.e., $H_{EB} = 0$. Interestingly, the largest EBs seem to occur at the transitions separating regimes where the loops are rectangular, with 'triangular tails', and 'double-triangles', respectively. While the values of $H_s^{np}$ and $H_s^{pn}$ in Fig.3(b) and $H_s^{av}$ in Fig.3(c) obtained at $83° < \theta < 86.5°$ look to be random, the corresponding loops are highly reproducible, as demonstrated in Fig.S4 in the Supplemental Material [30], which presents comparisons of two measurement runs for loops obtained at three particular angles (denoted by A, B, C in Fig.3(c)) where the $H_s^{av}$ unexpectedly deviates from the overall increasing trend, the EB is the largest, and the $H_s^{av}$ deviates from the overall decreasing trend, respectively.

The switching of spins in an ordinary ferromagnet is often caused by nucleation and motion of domain walls separating regions of opposite uniform magnetizations. The values of the associated switching fields depend on the angle $\theta$ between the magnetic field and the magnetization easy axis, following the Kondorsky relation [38-40]:

$$H_s(\theta) = H_s(0°)/\cos\theta \qquad (1)$$

As presented in Fig.3 as solid lines, Eq.1 describes $H_s^{np}$ and $H_s^{pn}$ as well as $H_s^{av}$ very well for $\theta$ up to ~80° while deviating from the experimental data as the magnetic field enters the angle regime where $R_{xy}(H)$ loops show EB and bow-tie-like structures. Nearly the same behavior occurs at a lower temperature $T$ = 50 K, as demonstrated in Fig.4(a)&(c) as well as Fig.S5 in the Supplemental Material [30]. At our lowest experimentally accessible temperature $T$ = 3 K, EB can be more unambiguously identified in a wider angle range (Fig.4(b)&(d)). However, this extension of the EB angle range arises from the exchange-biased rectangular loops (see Fig.S6 in the Supplemental Material [30]) while the angle regime (84° < $\theta$ < 88°) for the occurrence of bow-tie-like structures seems to be temperature-insensitive, as demonstrated by the large EB values at $\theta$ between 84° and 85° at all three temperatures.

Following the explanation proposed in Ref.15 and discussed above on the origin of the bow-tie-like structures, it seems reasonable to interpret the deviation from Eq.1 as a transition from a pure FM state to a mixture of FM/AFM phases, since a magnetic field in the vicinity of the *ab* plane should suppress the *c*-axis FM interaction while enhancing the in-plane AFM interaction. Recent neutron scattering experiments [25] reveal that AFM structures at $T > T_A$ probably originate from a temperature-induced local lattice instability due to geometrical frustration in the Co kagomé lattice. It is rational to hypothesize that such a local lattice instability can be enhanced by a magnetic field not parallel to the *c*-axis that is the magnetization easy axis, since a non-zero torque exerting on the Co ions can distort the Co kagomé lattice, thereby lowering the temperature at which AFM structures can exist. With increasing $\theta$, the torque becomes larger, resulting in stronger local lattice instability. Thus, tilting the magnetic field towards the *ab* plane at a fixed temperature can lead to similar effects as the temperature approaches $T_c$ at **H**||*c* on magnetic phases.

In summary, we investigated magnetic field orientation effects on the magnetism in the newly confirmed ferromagnetic Weyl semimetal $Co_3Sn_2S_2$. We found that tilting the magnetic field from the *c*-axis towards the *ab* plane at a fixed temperature can induce a change in the magnetic phases similar to that which occurs when the temperature is swept up to the Curie temperature while keeping the magnetic field aligned along the *c*-axis, as referred from the change (from rectangle to bow-tie-like structure) in the shape of the $R_{xy}(H)$ hysteresis loops. Furthermore, the bow-tie-like $R_{xy}(H)$ hysteresis loops observed near the *ab* plane can exhibit exchange bias, underscoring the significance of the AFM interaction for the magnetism in $Co_3Sn_2S_2$. Our results demonstrate a new way to tune the magnetic phases of $Co_3Sn_2S_2$ and shed light on the origin of the bow-tie-like structures in the hysteresis loops measured in the magnetic field along the *c*-axis.

**Acknowledgments**

We thank Vitalii Vlasko-Vlasov for stimulating discussions. Experimental design, transport measurements were supported by the U.S. Department of Energy, Office of Science, Basic Energy Sciences, Materials Sciences and Engineering. S.E.P & Z.L.X acknowledge support from the National Science Foundation grant# DMR-1901843. B.W. & B.S. were supported by the National Natural Science Foundation of China (NSFC) grant # U213010013, Natural Science Foundation of Guangdong Province grant# 2022A1515010035, Guangzhou, Basic and Applied Basic Research Foundation grant# 202201011798, and the open research fund of Songshan Lake materials Laboratory grant # 2021SLABFN11. Use of the Center for Nanoscale Materials, an Office of Science user facility, was supported by the U.S. Department of Energy, Office of Science, Office of Basic Energy Sciences, under Contract No. DE-AC02-06CH11357.

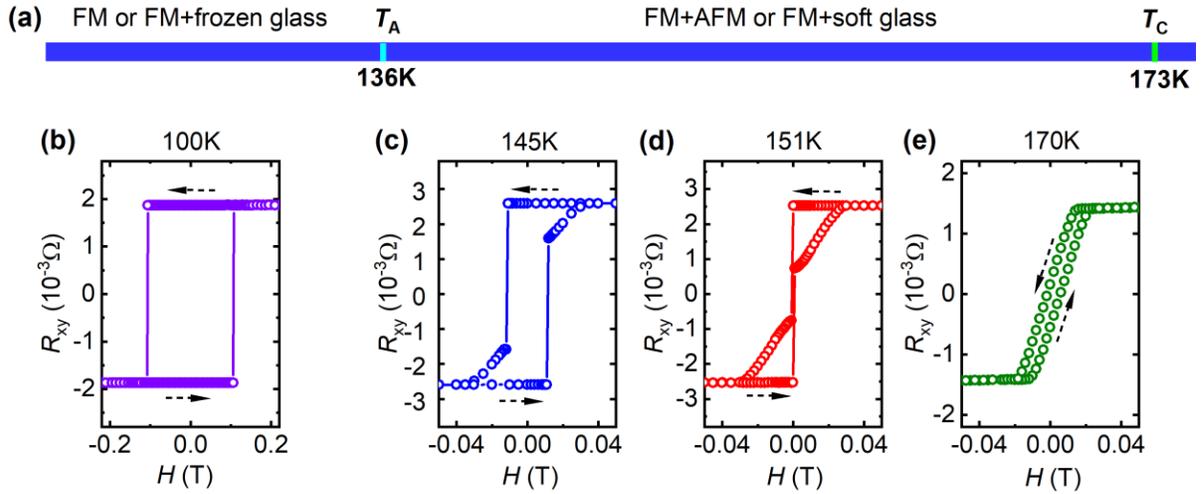

**Fig.1.** (a) Horizontal blue bar indicating a temperature $T_A$ separating various magnetic phases, with FM at $T < T_A$ and FM+AFM at $T > T_A$ revealed in μSR measurements in Ref.14 as well as FM+frozen glass at $T < T_A$ and FM+soft glass at $T > T_A$ suggested in Ref.15 from the shapes of $M(H)$ and $R_{xy}(H)$ hysteresis loops. (b)-(e) Representative $R_{xy}(H)$ hysteresis loops obtained at $T$ = 100 K, 145K, 151K and 170 K, respectively. The magnetic field was orientated along the *c*-axis. The sample has $T_c$ = 173 K (see Fig.S2 in the Supplemental Material [30]) and $T_A$ = 136 K (see FigS3 in the Supplemental Material [30]). The dashed arrows indicate the sweeping directions of the magnetic field.

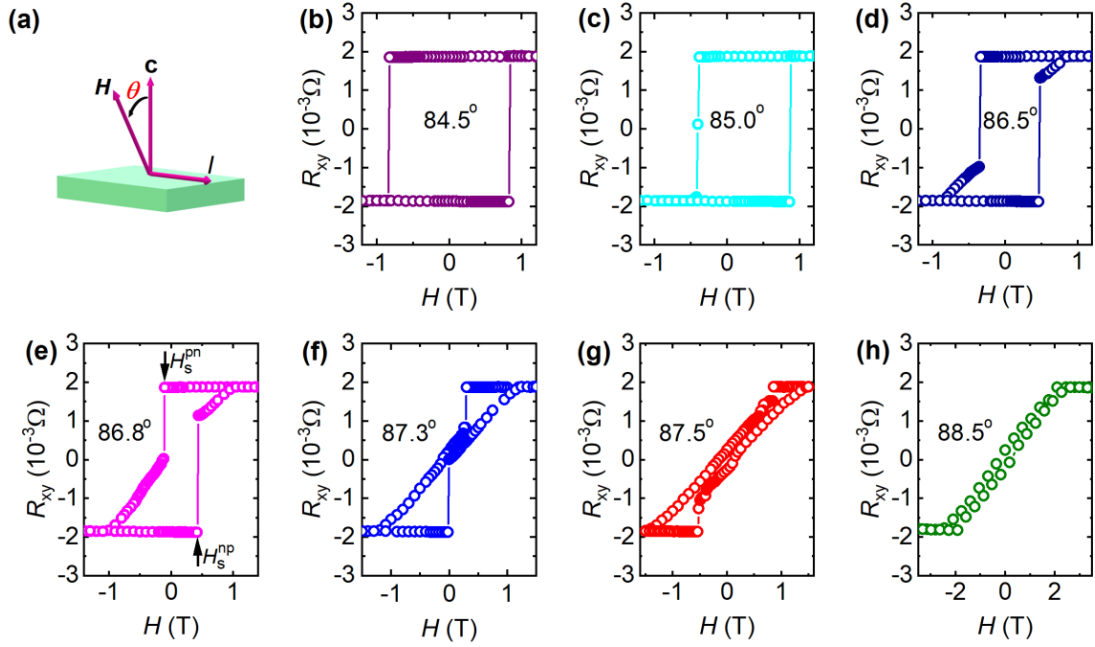

**Fig.2.** (a) Schematic of the magnetic field orientation, where $\theta$ is the angle between the applied magnetic field and the crystal's *c*-axis. (b)-(h) Representative $R_{xy}(H)$ hysteresis loops obtained at $\theta$ = 84.5°, 85.0°, 86.5°, 86.8°, 87.3°, 87.5°, and 88.5°, respectively. The data were obtained at *T* = 100 K with the magnetic field rotation plane perpendicular to the current flow direction. The definitions of $H_s^{pn}$ and $H_s^{np}$ are given in (e).

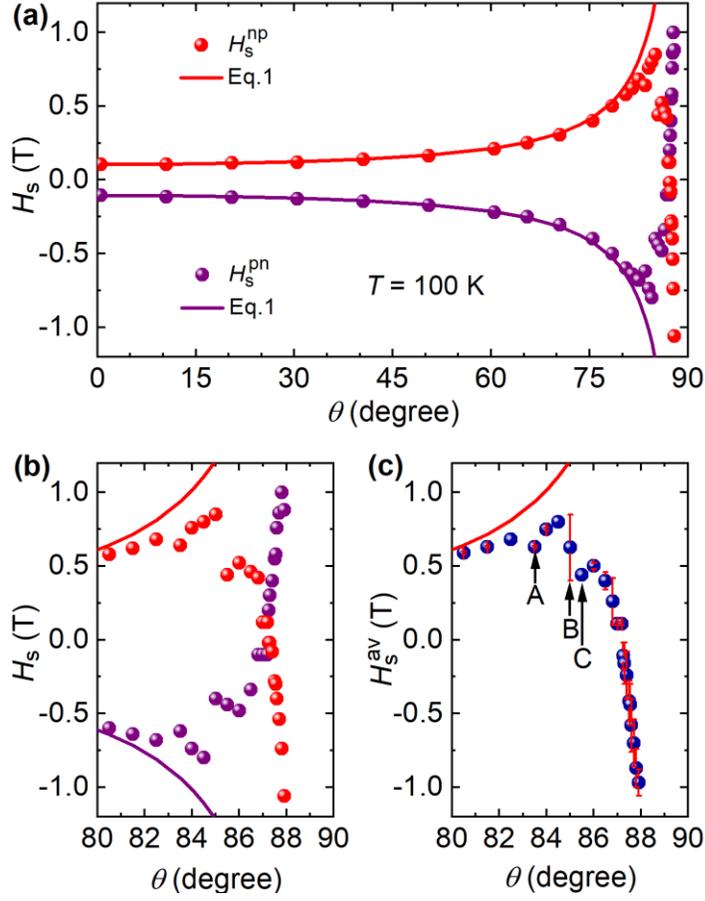

**Fig.3.** (a) Angle dependence of the switching fields $H_s^{pn}$ and $H_s^{np}$ (defined in Fig.2(e)). Symbols and lines represent experimental data and fits from Eq.1, respectively. (b) Expanded view of the data in (a) for the magnetic field orientated in the vicinity of the *ab* plane. (c) Angle dependence of the averaged switching field $H_s^{av} = (H_s^{np} - H_s^{pn})/2$ and the exchange bias $H_{EB} = (H_s^{np} + H_s^{pn})/2$, which are presented as solid symbols and error bars, respectively. Letters A, B and C denote the three angles at which the measurements were repeated twice for comparison and the data are presented in Fig.S6 in the Supplemental Material [30].

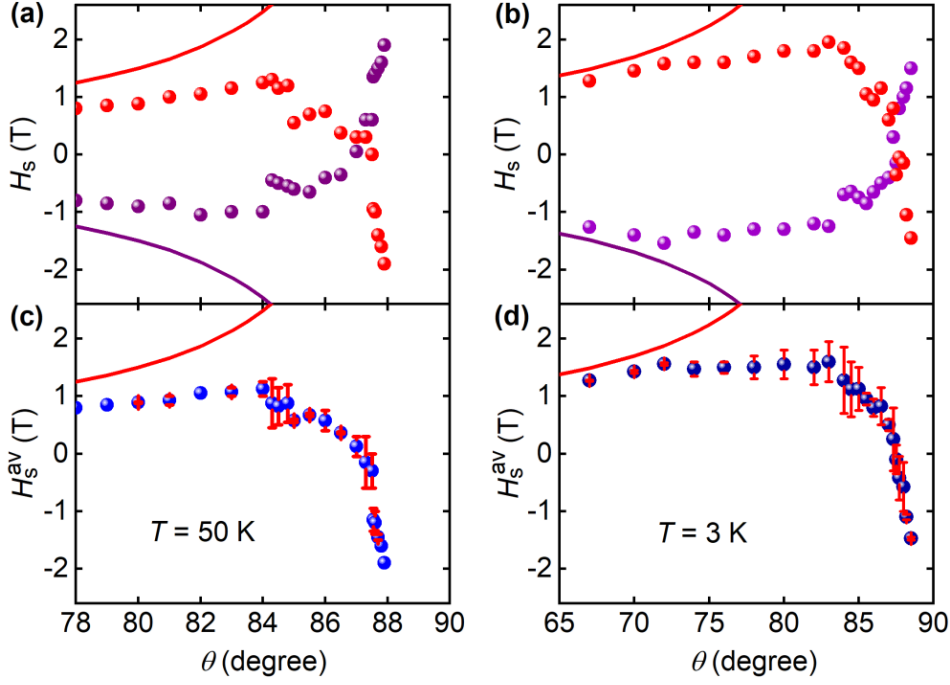

**Fig.4.** (a) and (b) Angle dependence of the switching fields $H_s^{pn}$ (purple symbols) and $H_s^{np}$ (red symbols) for $T = 50$ K and 3 K, respectively. Symbols and lines represent experimental data and fits from Eq.1, respectively. (c) and (d) Angle dependence of the averaged switching field $H_s^{av} = (H_s^{np} - H_s^{pn})/2$ and the exchange bias $H_{EB} = (H_s^{np} + H_s^{pn})/2$, which are presented as blue symbols and red error bars, respectively. For clarity we present only data in the angle regime where EB and bow-tie-like structures in the hysteresis loops are observed. Complete data sets of $H_s^{pn}$ and $H_s^{np}$ are presented in Fig.S5 in the Supplemental Material [30].

**Field Orientation Dependent Magnetic Phases In Weyl Semimetal Co$_3$Sn$_2$S$_2$**

By Samuel E. Pate et al.

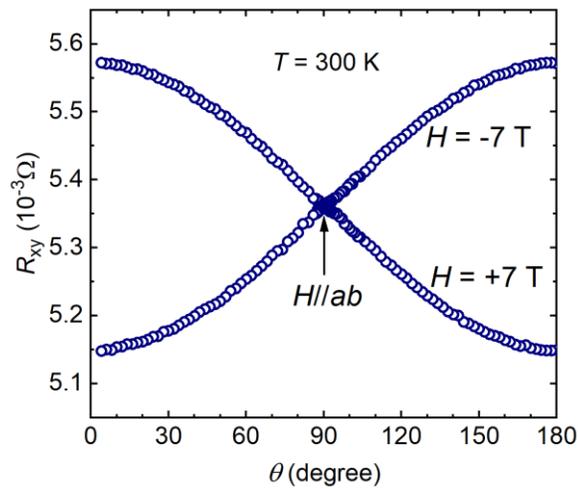

**Fig.S1**. Determination of the *ab* plane. Hall resistance versus angle $R_{xy}(\theta)$ curves were taken at $H = +7$ T and $-7$ T. The angle at which the two curves cross each other is defined as $\theta = 90°$, corresponding to the *ab* plane. Data were taken at $T = 300$ K

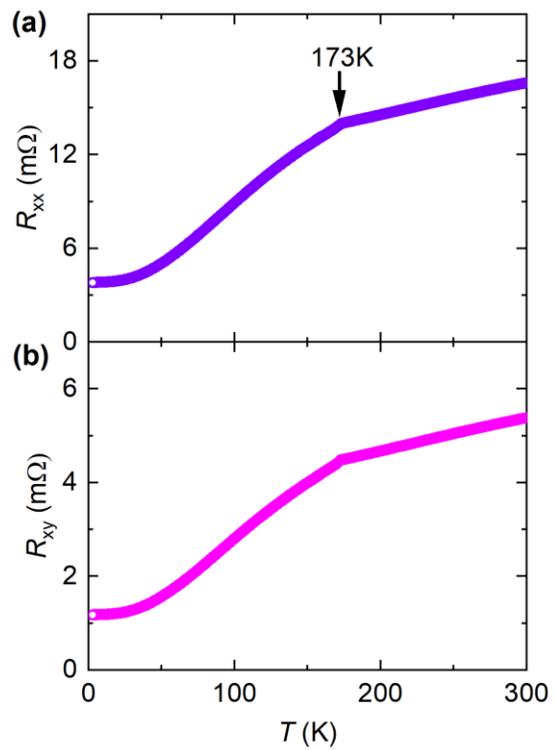

**Fig.S2.** (a) and (b) Temperature dependence of the longitudinal resistance $R_{xx}$ and the Hall resistance $R_{xy}$, respectively. Data were taken when the sample is cooled at zero field. The Curie temperature of $T_c$ = 173 K is indicated in (a).

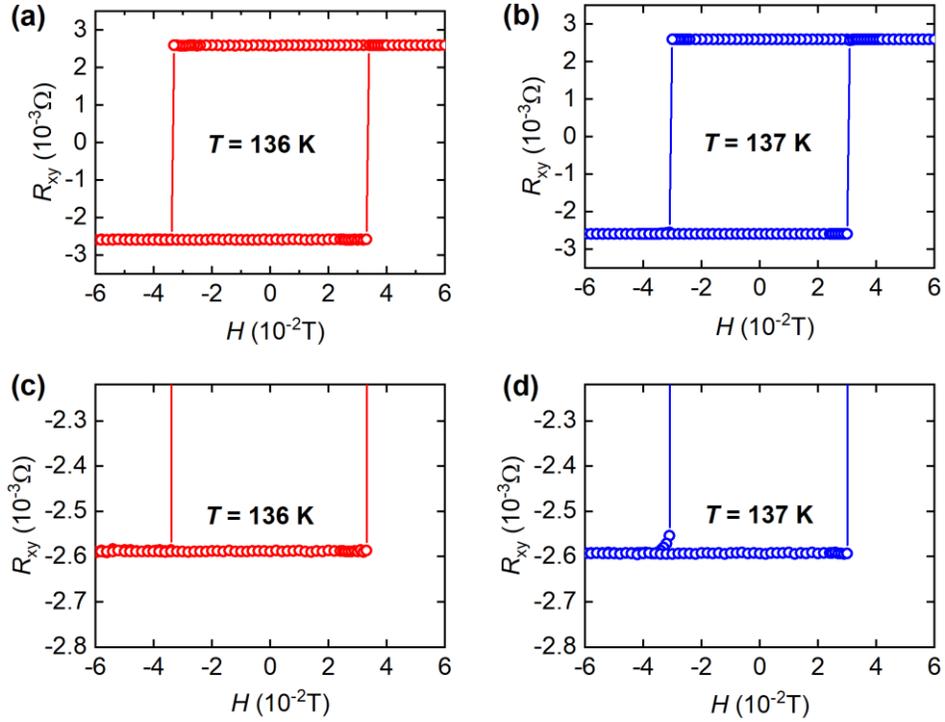

**Fig.S3.** (a) and (b) $R_{xy}(H)$ hysteresis loops obtained at $T = 136$ K and 137K. (c) and (d) Expanded views of the $R_{xy}(H)$ hysteresis loops in (a) and (b), respectively, which reveals $T_A = 136$ K.

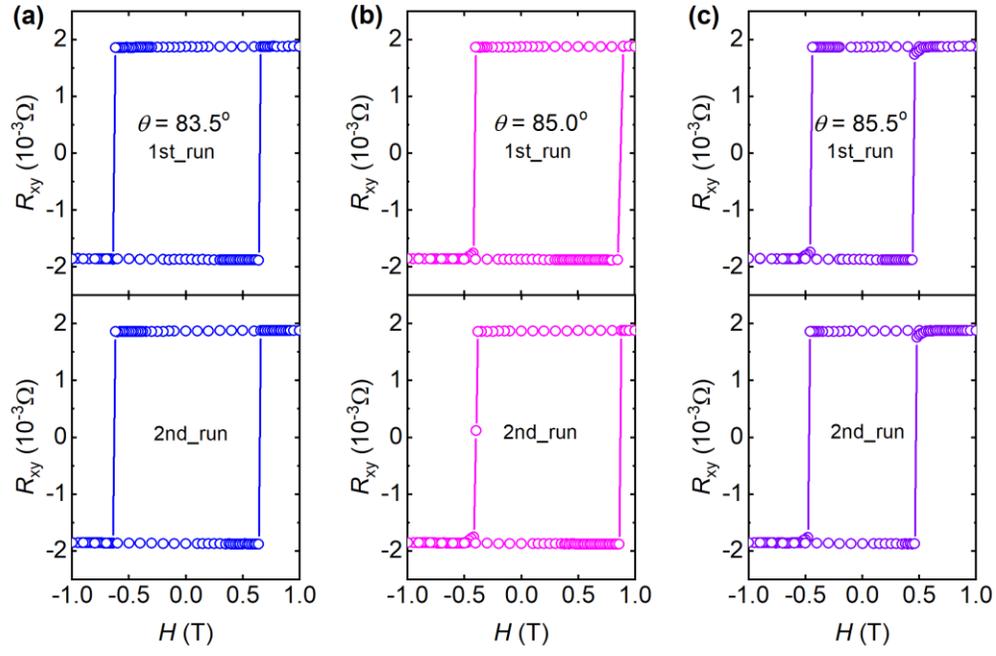

**Fig.S4.** (a), (b) and (c) Comparisons of two sets of $R_{xy}(H)$ loop measurements obtained at $\theta$ = 83.5°, 85.0° and 85.5° (denoted by A, B, C in Fig.3(c)), where the $H_s^{av}$ deviates from the overall increasing trend (A), the EB is the largest (B), and the $H_s^{av}$ deviates from the overall decreasing trend (C).

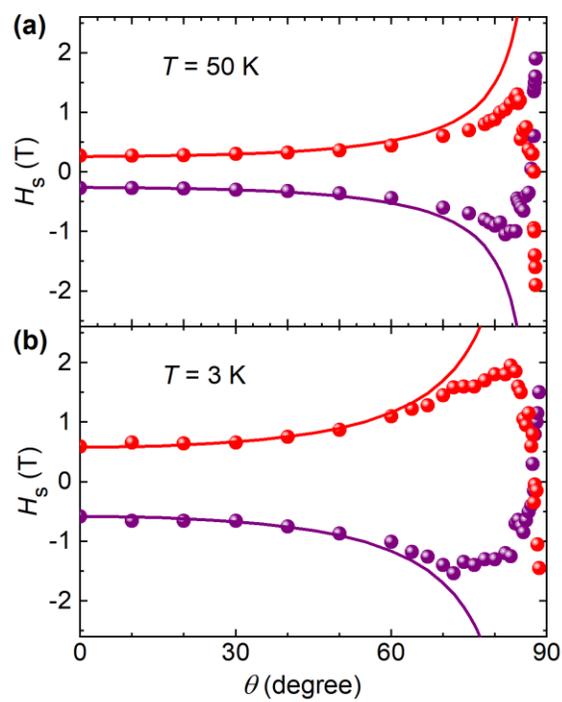

**Fig.S5**. (a) and (b) Angle dependence of the switching fields $H_s^{pn}$ (purple symbols) and $H_s^{np}$ (red symbols) for $T$ = 50 K and 3 K. Symbols and lines represent experimental data and fits from Eq.1, respectively.

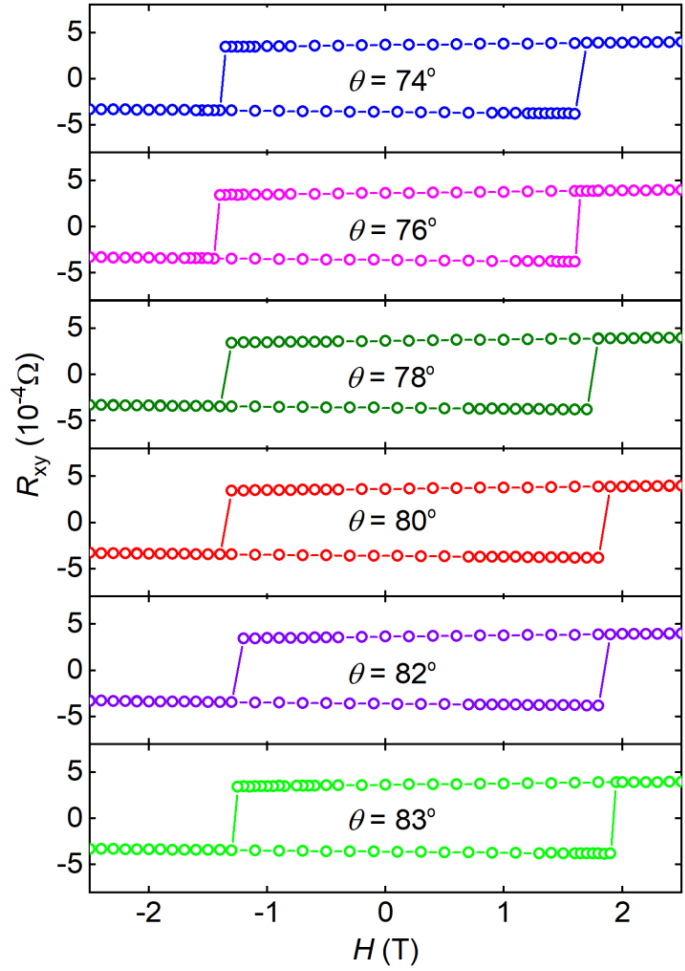

**Fig.S6.** $R_{xy}(H)$ loops obtained at $T = 3$ K and $\theta = 74°$, $76°$, $78°$, $80°$, $82°$ and $83°$. They are exchange-biased while showing no bow-tie-like structures. The values of $H_s^{pn}$, $H_s^{np}$ and $H_s^{av}$ deviate from the overall increase trend while being not as sensitive to the angle as those of the bow-tie-like $R_{xy}(H)$ loops (refer to Fig.4(b) and 4(d)).